\documentclass[aps,prb,amsmath,showpacs,twocolumn,superscriptaddress]{revtex4}

\begin{document}
\title{Non-abelian Berry's phase and Chern numbers in higher spin pairing condensates}
\author{Chyh-Hong Chern}
\affiliation{Department of Physics, McCullough Building, Stanford
University, Stanford  CA~~94305-4045}
\author{Han-Dong Chen}
\affiliation{Department of Applied Physics, McCullough Building,
Stanford University, Stanford CA 94305-4045}
\author{Congjun Wu}
\affiliation{Department of Physics, McCullough Building, Stanford
University, Stanford  CA~~94305-4045}
\author{Jiang-Ping Hu}
\affiliation{Department of Physics
and Astronomy, University of California, Los Angels, CA
90095-1547}
\author{Shou-Cheng Zhang}
\affiliation{Department of Physics, McCullough Building, Stanford
University, Stanford  CA~~94305-4045}

\begin{abstract}
We show that the non-Abelian Berry phase emerges naturally in the
s-wave and spin quintet pairing channel of spin-$3/2$ fermions.
The topological structure of this pairing condensate is
characterized by the second Chern number. This topological
structure can be realized in ultra-cold atomic systems and in
solid state systems with at least two Kramers doublets.
\end{abstract}

\pacs{74.90.+n, 71.10.Fd, 03.75.Nt, 02.40.Re}
\maketitle
\section{Introduction}
Topological gauge structure and Berry's phase\cite{BERRY1984} play
an increasingly important role in condensed matter physics. The
quantized Hall conductance can be deeply understood in terms of
the first Chern class\cite{THOULESS1982,KOHMOTO1985}. The
fractional quantum Hall effect can be fundamentally described by
the $U(1)$ topological Chern-Simons gauge theory\cite{ZHANG1989}.
The effective action for ferromagnets\cite{Auerbach} and
one dimensional antiferromagnets\cite{Haldane1983,Haldane1983A} contain Berry
phase terms which fundamentally determine the low energy dynamics.
More recently, Berry's phase associated with the BCS
quasi-particles in pairing condensates has also been studied
extensively. \cite{volovik,Read2000,Hatsugai2003}.

However, most of the well known applications of the geometrical
phase in condensed matter physics involves Berry's\cite{BERRY1984}
original abelian U(1) phase factor. More recently, the non-abelian
SU(2)\cite{WILCZEK1984,ZEE1988,AVRON1988} Berry's phase (or
holonomy, to be precise) has been systematically investigated in
the context of condensed matter systems. Demler and
Zhang\cite{DEMLER1999} investigated the quasi-particle wave
functions in the unified SO(5) theory of antiferromagnetism and
superconductivity, and found that the SDW and the BCS
quasi-particle states accumulate an SU(2) Berry's phase (or
holonomy) when the order parameter returns to itself after an
adiabatic circuit. Zhang and Hu\cite{ZHANG2001} found a higher
dimensional generalization of the quantum Hall effect based on a
topologically non-trivial SU(2) background gauge field. Rather
surprisingly, the non-abelian SU(2) holonomy also found its deep
application in the technologically relevant field of quantum
spintronics\cite{MURAKAMI2003,Murakami2003A}. All these condensed matter
applications are underpinned by a common mathematical framework,
which naturally generalizes the concept of Berry's U(1) phase
factor. This new class of applications is topologically
characterized by the second Chern class and the second Hopf map,
and applies to fermionic systems with time reversal invariance.

In this paper, we investigate the non-trivial topological
structures associated with the higher spin condensates. We first
review the momentum space gauge structure of the spin one
condensate, namely the A phase of $^3He$. As it has been
pointed out\cite{volovik}, the momentum space gauge structure of the pairing
condensate is given by that of the t'Hooft-Polyakov monopole. We
then investigate the new system of spin 2 (quintet) pairing
condensate of the underlying spin 3/2 fermions. The most general
Hubbard model of spin 3/2 fermions has recently been introduced
and investigated extensively by Wu et al.\cite{wu2003}, who found
that the model always has a generic SO(5) symmetry in the spin
sector. Building on this work, we show here that the fermionic
quasi-particles of the quintet pairing condensate can be described
by the second Hopf map. Similar to the SDW+BCS system investigated
by Demler and Zhang\cite{DEMLER1999}, the quasi-particles of the
quintet pairing condensate also accumulate an SU(2) holonomy. The
quintet pairing condensate can be experimentally realized in a
number of systems. Cold atoms with spin 3/2 in the continuum or on
the optical lattice can be accurately described by the model of
local contact interactions\cite{wu2003} $U_0$ and $U_2$. These
interaction parameters can be experimentally tunes over a wide
range, including the range for stable quintet condensates.
Effective spin 3/2 fermions can also be realized in solid state
systems with at least two Kramers doublets, for example in bands
formed by $P_{3/2}$ orbitals.

In the rest of this paper, we shall use spin-1/2 system to be
short for the spin-$1/2$ superfluid $^3He$-A and spin-$3/2$ system
for the $s$-wave spin-3/2 superconductor in the quintet channel.
The repeated indices are summed assumably throughout this paper.

\section{superfluid $^3He$-A}\label{SECTION-He3}
\subsection{Goldstone manifold and the $1^{st}$ Hopf map}
The general form of the equilibrium order parameter in the $^3He$-A phase can be written as\cite{volovik}
\begin{eqnarray}
  \langle{\Delta}(k)_{ai}\rangle = \Delta_k \hat{d}_a
  \left(\hat{e}_i^{(1)}+i\hat{e}_i^{(2)}\right),
\end{eqnarray}
where the spin index ($a$) and and orbital index($i$) run from $1$ to $3$.
$\Delta_k$ is a complex number that contains the information of the magnitude and the $U(1)$ phase. $\hat{d}$ is the normal vector of the plane to which the
spin direction is restricted. The orthogonal vectors $\hat{e}^{(1)}$, $\hat{e}^{(2)}$ and $\hat{l}=\hat{e}^{(1)}\times \hat{e}^{(2)}$ form a local physical coordinate frame.

The Goldstone manifold of the order parameter is
given by\cite{volovik}
\begin{eqnarray}
  R_a&=&G/H=\frac{U(1)\times SO(3)^{(L)}\times SO(3)^{(S)}}
  {SO(2)^{(S)}\times U(1)^{combined}\times Z_2^{combined}}\nonumber\\
  &=&S^2\times SO(3)_{relative}/Z_2.
\end{eqnarray}
Here $SO(3)_{relative}$ denotes such rotations about
the axis $\hat{l}$ that lead to new degenerate states which are relative towards
gauge transformations\cite{volovik}.
  The $U(1)^{combined}$ comes from the fact that
the A-phase state is invariant under combined
transformation\cite{volovik} of the gauge transformation with the
parameter $\phi$ from the $U(1)$ group and the orbital rotation of
$\hat{e}^{(1)}$, $\hat{e}^{(2)}$ and $\hat{l}$ about axis
$\hat{l}$ by the same angle $\phi$. The $Z_2^{combined}$ denotes
the combined operation that $\hat{d}\rightarrow
-\hat{d},\Delta_k\rightarrow -\Delta_k$. This combined discrete
symmetry leads to the existence of half-quantum
vortices\cite{volovik}.
Around a half-quantum vortex, the vector field $\hat{d}$ is continuously rotated into
$-\hat{d}$ and the $U(1)$ phase of $\Delta_k$ continuously
evolves from $0$ to $\pi$ when the order parameter returns to itself after an adiabatic circuit.

If we fix
the local orthogonal frame in an arbitrary direction and
adiabatically move the quasi-particle around a line defect of
half-quantum vortex, the trajectory of the order parameter is a
closed loop on the $S^2/Z_2$ space. On the other hand,
 the degrees of freedom of the quasi-particle (a 2-dimensional spinor)
forms a 3-dimensional sphere $S^3$ and the trajectory of the quasi-particle on $S^3$ is not closed. This adiabatic evolution
defines the following map
\begin{eqnarray}
S^3{\longrightarrow} S^2/Z_2. \label{1st hopf}
\end{eqnarray}
In the topological terminology, Eq.(\ref{1st hopf}) is determined
by the third homopotic group denoted by $\pi_3(S^2/Z_2)$.  Due to
a theorem in the Homopoty theory\cite{Hilton1953}
\begin{eqnarray}
\pi_k(S^n/Z_2)=\pi_k(S^n),\ \textrm{for}\ k\geq 2.
\end{eqnarray}
Eq.(\ref{1st hopf}) is homopotically equivalent to the first Hopf
map $S^3\rightarrow S^2$ {\it i.e.}, U(1) Berry phase in the FQHE
and other nano-structure in the semiconductors\cite{Yau2002,
yang2002}.
\subsection{Berry connection, $1^{st}$ Chern number, t'Hooft-Polyakov monopole and Dirac monopole}
If we define the spinor as
 \begin{eqnarray}
    \Psi_{k}^\dag=
  \left(c^\dag_{k,\frac{1}{2}},c^\dag_{k,\frac{-1}{2}},
 c_{-k,\frac{1}{2}}, c_{-k,\frac{-1}{2}}\right),
\end{eqnarray}
the mean field Hamiltonian for $^3He$-A is given by
\begin{eqnarray}
    \mathcal{H}=\sum_k^{'}\Psi^\dag_kH_k\Psi_k
\end{eqnarray}
with
\begin{eqnarray}
  H_k = \left(
  \begin{array}{cc}
    \epsilon_k\sigma^0 & {\bf \Delta}_k\\ ~ \\
    {\bf \Delta}_k^\dag  & -\epsilon_k\sigma^0
  \end{array}
  \right), \label{He-Himiltonian}
\end{eqnarray}
where ${\bf \Delta}_k=-\Delta_k \hat{d}_a\sigma^a R$ and $R=-i\sigma^2$. $\epsilon_k$ is the kinetic energy on the lattices referenced from the Fermi surface and the summation of momentum $k$ is over half of the Brillouin zone to avoid the double counting. Here, $\sigma^0$ is the $2\times 2$ identity matrix and $\sigma^{1,2,3}$ are Pauli matrices.

The Berry phase connection (BPC) is defined by the differential
change of states projecting to themselves.  In this paper, it will
be illustrated by using the state with a positive eigenvalue. BPC
obtained from the state with a negative eigenvalue is simply the
complex conjugate to the one with the eigenvalue of a different
sign. The BPC and its field strength can be obtained respectively.
\begin{eqnarray}
A_a=-iA^c_a\frac{\sigma_c}{2}, \quad
A_a^c=\varepsilon_{abc}\frac{\hat{d}_b}{d}. \label{so3_potential}
\end{eqnarray}
and
\begin{eqnarray}
F_{bc}^a=&&\partial_bA_c^a-\partial_bA_c^a+\epsilon_{ade}A^d_bA^e_c
\nonumber \\=&&
-\frac{1}{d^2}\epsilon_{bce}\hat{d}_e\hat{d}_a,\label{so3_field}
\end{eqnarray}
The gauge invariant magnetic field can be defined as
\begin{eqnarray}
B^a=\frac{1}{2}\epsilon_{abc}F_{bc}^ed_e=-\frac{\hat{d}_a}{d^2}.
\end{eqnarray}
This is a U(1) magnetic-monopole like field in the $d$-space.  It
emerges when there are line defects in the $\hat{d}$ field,
$e.g.$, half-quantum vortices in the superfluid $He-3A$ phase. If
we transport the spin-1/2 fermion adiabatically around the
vortex,the electronic wavefunction gains the phase accumulated due
to the $\hat{d}$ field, as we discussed previously.  Moreover, the
first Chern number can be computed easily.
\begin{eqnarray}
\mathcal{C}_1=\frac{1}{4\pi}\oint\vec{B}\cdot \vec{d}S=-1.
\end{eqnarray}

This is the famous t'Hooft-Polyakov monopole (TPM) \cite{volovik,
Weinberg}. Different from the Dirac monopole, the gauge field of
TPM is non-abelian and finite everywhere over $S^2/Z_2$ while the
Dirac magnetic monopole is abelian and has a singularity string.
There is a deep and direct relation between them, which can be
achieved by a singular gauge
transformation\cite{Boulware1976,Ohnuki1993, chern2004}.

The present $SO(3)$ Berry phase defines a $SO(3)$ gauge theory on
$S^2/Z_2$.  Using the covariant derivative $D_a=\partial_a+A_a$,
the $SO(3)$ generators in the presence of t'Hooft-Polyakov
monopole can be written as\cite{yang1978}
\begin{eqnarray}
L_{ab}=\Lambda_{ab}-id^2f_{ab}, \ a=1,2,3 \label{generator}
\end{eqnarray}
where $\Lambda_{ab}=-id_aD_b+id_bD_a$ and
$f_{ab}=-iF^c_{ab}\frac{\sigma_c}{2}$. Defining
$I_{a}=\frac{1}{2}\epsilon_{abc}L_{bc}$, one finds easily that
$[I_a, I_b]=i\epsilon_{abc}I_c$ satisfying the $SO(3)$ algebra.
Using Eq.(\ref{so3_potential}) and Eq.(\ref{so3_field}), one can
show
\begin{eqnarray}
L_{ab}=L^{(0)}_{ab}+\epsilon_{abc}\frac{\sigma_c}{2}
\end{eqnarray}
where $L^{(0)}$ is the orbital angular momentum, defined by
$L^{(0)}_{ab}=-id_a\partial_b+id_b\partial_a$.
Define\cite{Ohnuki1993}
\begin{eqnarray}
    V=\exp\left[i\frac{\vartheta\sigma_{3a}d_a}
    {\sqrt{d^2-d^2_3}}\right]
\end{eqnarray}
where $\sigma_{ab}=\epsilon_{abc}\sigma_c/2$ and
$\cos\vartheta=d_3/d$.  One can perform a singular $SO(3)$ gauge
transformation on $L_{ab}$ such that $J_{ab}=VL_{ab}V^\dag$ where
\begin{eqnarray}
&&J_{\mu \nu}=-id_\mu\partial_\nu+id_\nu\partial_\mu+\epsilon_{\mu
\nu}\frac{\sigma_3}{2} \\ &&J_{\mu
3}=-id_\mu\partial_3+id_3\partial_\mu-\epsilon_{\mu
\nu}\frac{d_\nu}{d+d_3}\frac{\sigma_3}{2}
\end{eqnarray}
with $\mu,\nu=1,2$. $\epsilon_{\mu\nu}$ is the antisymmetry tensor
which has only one component $\epsilon_{12}=1$ in this case.  It
is obvious that $J_{\mu\nu}=J_{12}$ forms the $U(1)$ generator on
$S^2/Z_2$.  From the definition of Eq.(\ref{generator}), one can
extract the $U(1)$ BPC regardless the unnecessary
$\frac{\sigma_3}{2}$.
\begin{eqnarray}
a_\mu=-i\epsilon_{\mu \nu}\frac{d_\nu}{d(d+d_3)}, \ a_3=0
\end{eqnarray}
and the finite $U(1)$ field strength over $S^2/Z_2$
\begin{eqnarray}
F_{ab}=i\epsilon_{abc}\frac{d_c}{d^3}
\end{eqnarray}

We should notice that the singular gauge transformation we used
has a singularity string along the negative $z$-axis. Therefore,
while the covariant $SO(3)$ BPC is finite over the whole $\hat{d}$
field, the $U(1)$ BPC has a singularity string which is reflected
through the transformation.  This transformation is only valid on
the northern hemisphere including the equator of the $S^2/Z_2$.
One is able to choose another gauge which has the singularity
along the positive $z$-axis to describe the transformation on the
southern hemisphere.

The role of this singular gauge transformation is very intriguing.
We can view the covariant $SO(3)$ gauge potential $A^c_a$ in
Eq.(\ref{so3_potential}) as a vector $d_c$ pointing in the isospin
space.  The singular gauge transformation is nothing but the
rotation of the spin vector from $d_c$ to $d_3$.  Therefore, the
invariant subgroup of $SO(3)$ is emerged from the isometry group
of the equator of $S^2/Z_2$, which is $U(1)$.  This mechanism
accounts for the appearance of the $U(1)$ Berry phase in this
problem.  As a result, the $SO(3)$ Berry phase in this system is
essentially equivalent to the $U(1)$ Berry phase.

\section{$s$-wave quintet pairing condensate in spin-3/2 system}\label{SECTION-spin-3/2}
\subsection{Goldstein manifold and the $2^{nd}$ Hopf map}
Another candidate for non-trivial gauge structures is the spin-3/2
fermionic system with contact interaction, in which an exact
$SO(5)$ symmetry was identified recently\cite{wu2003}. It may be
studied in the ultra-cold atomic systems, like $^9$Be, $^{132}$Cs,
$^{135}$Ba, $^{137}$Ba. The four-component spinor
$(c_{\frac{3}{2}},c_{\frac{1}{2}},
c_{\frac{-1}{2}}, c_{\frac{-3}{2}})^T$ forms the spinor
representations of the $SU(4)$ group which is the unitary
transformation of the four-component complex spinor. The kinetic
energy term has the explicit $SU(4)$ symmetry. However, the
$s$-wave contact interaction term breaks the $SU(4)$ symmetry to
$SO(5)$.
 Because of the $s$-wave scattering, there are only the
singlet and quintet channels as required by the Pauli's exclusion
principle.  Interestingly, the spin $SU(2)$ singlet and quintet
channels interaction can also be interpreted as $SO(5)$ group's
singlet and 5-vector representations.

The Cooper pair structures has also been studied in spin-3/2
system\cite{ho1999, wu2003}.  The singlet and quintet pairing
channel operators are described by
\begin{eqnarray}
\eta^\dagger(r)&=&\frac{1}{2} c^\dagger_\alpha(r) R_{\alpha\beta}
c^\dagger_\beta(r)\nonumber \\
\chi_a^\dagger(r)&=&\frac{-i}{2} c^\dagger_\alpha(r)
(\Gamma^a R)_{\alpha\beta} c^\dagger_\beta(r)\nonumber.
\end{eqnarray}
where $\Gamma^a$ are the $SO(5)$ Gamma matrices which takes the
form
\begin{eqnarray}
 \Gamma^1\!=\! \left( \begin{array}{cc}
               0          & iI  \\
               -iI          & 0            \end{array} \right), \
\Gamma^{i} \!=\! \left(
\begin{array}{cc}
                  \sigma_i & 0  \\
               0&- \sigma_i   \end{array} \right), \
 \Gamma^5 \!=\! \left( \begin{array}{cc}
                0         & -I  \\
                I         & 0            \end{array} \right) &&
                \label{so5 gamma matrix}
\end{eqnarray}
satisfying the Clifford algebra
$\{\Gamma^a,\Gamma^b\}=2\delta^{ab}$. The $SO(5)$ charge conjugate
matrix $R$ is given by
\begin{eqnarray}
R=-i\left( \begin{array}{cc} \sigma_2 & 0 \\ 0 & \sigma_2
\end{array} \right)
\end{eqnarray}
The quintet pairing structure is spanned by the five polar like
operators $\chi^\dagger_{1\sim5}$, whose expectation value has a
5-vector and a phase structure as $ d_a e^{i\phi}$.
The Goldstone
manifold for the quintet pairing is
\begin{eqnarray}
  R_{3/2}&=& \frac{SO(5)_s\otimes SO(3)_L\otimes U(1)}
  {SO(4)_s\otimes SO(3)_L \otimes Z_2}\nonumber\\
  &=&S^4\otimes U(1)/Z_2,
\end{eqnarray}
where the $Z_2$ symmetry comes from the combined
operations $d_a\rightarrow -d_a, \phi\rightarrow \phi+\pi$.

Because the 4-component spinor forms the 7-dimensional sphere,
similar to the spin $1/2$ case, the adiabatic transportation of
the quasi-particle around a half-quantum vortex in our spin-3/2 system defines
a map
\begin{eqnarray}
S^7{\longrightarrow} S^4/Z_2,
\end{eqnarray}
which is homopotically equivalent to the second Hopf map $i.e.$
$S^7\rightarrow S^4$.

\subsection{Berry connection, $2^{nd}$ Chern number, SO(4) monopole and Yang monopole}
Let us introduce the spinor
\begin{widetext}
\begin{eqnarray}
    \Psi_{k}^\dag=
  \left(c^\dag_{k,\frac{3}{2}},c^\dag_{k,\frac{1}{2}},
    c^\dag_{k,\frac{-1}{2}},c^\dag_{k,\frac{-3}{2}},
    c_{-k,\frac{3}{2}},c_{-k,\frac{1}{2}},
    c_{-k,\frac{-1}{2}},c_{-k,\frac{-3}{2}}\right),
\end{eqnarray}
\end{widetext}
where $c^\dag_{k\sigma}$
is the creation operator of an electron with the spin component
$\sigma$ and momentum $k$. The mean field Hamiltonian can be written as
\begin{eqnarray}
    \mathcal{H}=\sum_k^{'}\Psi^\dag_kH_k\Psi_k
\end{eqnarray}
with
\begin{eqnarray}
  H_k = \left(
  \begin{array}{cc}
    \epsilon_k\Gamma^0 & {\bf \Delta}_k\\ ~ \\
    {\bf \Delta}_k^\dag & -\epsilon_k\Gamma^0
  \end{array}
  \right) \label{grand Himiltonian}
\end{eqnarray}
where $\epsilon_k$ is the kinetic energy on the lattices
referenced from the Fermi surface.  ${\bf \Delta}_k= -\Delta_k d_a
\Gamma^a R$ while $\Delta_k$ contains the magnitude and the phase
of the superconducting (SC) order parameter. The momentum $k$ is
summed only over half of the Brillouin zone to avoid the double
counting. The subscript $a$ runs from $1$ to $5$.  $d_a$ forms a
4-dimensional sphere $S^4$. $\Gamma^0$ is the $4\times 4$ identity
matrix, and $\Gamma^a$ are given by Eq.(\ref{so5 gamma matrix}).
The eigenvalues of Eq.(\ref{grand
Himiltonian}) are
\begin{eqnarray}
 \lambda = \pm E_k= \pm
\sqrt{\epsilon_k^2+|\Delta_k|^2}
\end{eqnarray}
 and their corresponding
engenvectors are
\begin{eqnarray}
\psi^+_\alpha(k)=\frac{1}{\sqrt{(E_k+\epsilon_k)^2+|\Delta|^2}}\left(
\begin{array}{c} (E_k+\epsilon_k)|\alpha\rangle \\ ~ \\ {\bf \Delta}_{\bf k}^\dag|\alpha\rangle \end{array} \right) && \label{so5 spinor}\\
\nonumber\\
\psi^-_\alpha(k)=\frac{1}{\sqrt{(E_k+\epsilon_k)^2+|\Delta|^2}}\left(
\begin{array}{c}{\bf \Delta}_{\bf k}|\alpha\rangle \\ ~ \\
(E_k+\epsilon_k)|\alpha\rangle
\end{array} \right) &&
\end{eqnarray}
where $|\alpha\rangle$ are $SU(4)$ spinors.

We are interested in the system with the presence of half-quantum
vortices.  The formation of this kind of vortices is very similar
to the ones in the spin-1/2 system.  If we transport the spin-3/2
fermion adiabatically around one of them, a nontrivial phase is
accumulated due to the $\hat{d}$ field.
The BPC and the covariant field strength can be obtained respectively by
\begin{eqnarray}
A_a = \frac{i}{d^2}d_c\Gamma^{ca}, \ a=1,2,3,4,5 \label{so5 gauge
potential}
\end{eqnarray}
and
\begin{eqnarray}
F_{abc}=-\frac{i}{d^3}(d_a\Gamma^{bc}+d_b\Gamma^{ca}+d_c\Gamma^{ab})
\end{eqnarray}
where $\Gamma^{ab}=\frac{1}{4i}[\Gamma^a, \Gamma^b]$ making up of
the $SO(5)$ generators.  Similar non-abelian gauge structures also
appeared in the pseudoparticle field in high
dimensions\cite{jackiw1976,adler1972,adler1973}.  Instead of the
first Chern number, we have the non-vanishing second Chern number.
\begin{eqnarray}
\mathcal{C}_2=-\frac{1}{96\pi^2}\oint d\Omega_d \text{Tr}
\left(F_{abc}F_{abc}\right)=1.
\end{eqnarray}
where $d\Omega_d$ denotes the integration over the angular part of
$d_a$.  The field strength on $S^4$, $f_{ab}=[D_a, D_b]=\partial_a
A_b -\partial_b A_a+[A_a, A_b]$, can be obtained
\begin{eqnarray}
f_{ab}=-i\frac{1}{d^2}P_{ab,cd}\Gamma^{cd}
\end{eqnarray}
where
$P_{ab,cd}=\frac{1}{2}(\delta_{ac}\delta_{bd}-\delta_{ad}\delta_{bc}+\delta_{ad}d_bd_c-\delta_{bd}d_ad_c-\delta_{ac}d_bd_d+\delta_{bc}d_ad_d)$.
$P_{ab,cd}$ is a $10\times 10$ matrix because $a$ and $b$ are
anti-symmetric as well as $c$ and $d$.  Similar to the projection
operator $\delta_{\mu\nu}-\hat{q}_\mu \hat{q}_\nu$ in QED,
$P_{ab,cd}$ is the transverse projection operator from
10-dimensional space to 6-dimensional space satisfying
$d_aP_{ab,cd}=0$. The relation between $F_{abc}$ and
ordinary field strength $f_{ab}$ can be revealed if we define
$G_{ab}$ which is dual to $F_{abc}$ by
$G_{ab}=\epsilon_{abcde}F_{abc}$\cite{adler1972, adler1973}. Then,
one can show
\begin{eqnarray}
f_{ab}=\frac{1}{2}\epsilon_{abcde} \frac{d_c}{d}G_{de}.
\end{eqnarray}
Because of the projection operator $P_{ab,cd}$ in $f_{ab}$, the
fundamental degrees of freedom of the gauge structure in this
problem is not $SO(5)$ but $SO(4)$, because SO(4) has six
generators.  We shall show that it is able to make a route from
$f_{ab}$ to the $SO(4)$ gauge field strength using a singular
gauge transformation. Similar to the analysis in the second
section, the transformation operator in this $SO(5)$ case has the
following form
\begin{eqnarray}
 U=\exp\left[i\frac{\vartheta\Gamma^{5\mu}d_\mu}
 {\sqrt{d^2-d^2_5}}\right], \quad \mu=1,2,3,4 \label{U}
\end{eqnarray}
where $\cos\vartheta=d_5/d$.
Then, Eq.(\ref{so5 gauge potential}) becomes
\begin{eqnarray}
&&a_\mu=\frac{-i}{d(d+d_5)}\Sigma^{\mu\nu}d_{\nu}, \quad
\mu=1,2,3,4\\
&&a_5=0,
\end{eqnarray}
where $\Sigma_{\mu\nu}$ are the $SO(4)$ generators in the
$(\frac{1}{2},0)\oplus(0,\frac{1}{2})$ representation which have
the following form
\begin{eqnarray}
\Sigma_{\mu\nu}=\left( \begin{array}{cc}
\eta^i_{\mu\nu}\frac{\sigma_i}{2} & 0 \\ ~\\ 0 &
\bar{\eta}^i_{\mu\nu}\frac{\sigma_i}{2} \end{array}\right)\label{sigular}
\end{eqnarray}
where
$\eta^i_{\mu\nu}=\epsilon_{i\mu\nu4}+\delta_{i\mu}\delta_{4\nu}-\delta_{i\nu}\delta_{4\mu}$
is the t'Hooft symbol, $\mu$ and $\nu$ runs from 1 to 4.  In this
reducible representation of $SO(4)$ gauge group, one can easily
distillate the $SU(2)$ ingredients because $SO(4)=SU(2)\otimes
SU(2)$.  The self-dual $SU(2)$ gauge field is given by
\begin{eqnarray} \nonumber
&&a_\mu=\frac{-i}{d(d+d_5)}\eta^i_{\mu\nu}d_{\nu}\frac{\sigma_i}{2}
\quad \mu=1,2,3,4\\
&&a_5=0.  \label{su2 gauge potential}
\end{eqnarray}
Similar to the spin-1/2 system, we obtain the $SO(4)$ BPC which is
only defined on the northern hemisphere with the equator.  The
singularity string along the negative $z$-axis inherits from the
singular gauge transformation.  The role of the singular gauge
transformation by $U$ can be also interpreted as the rotation of a
5-dimensional vector from an arbitrary direction $d_a$ to $d_5$ in
the 5-dimensional isospin space. Therefore, the invariant subgroup
is the isometry group of the equator $S^3/Z_3$, which is $SO(4)$.
Surprisingly, the the representation we achieve in SO(4) gauge
theory is the reducible $(\frac{1}{2},0)\oplus(0,\frac{1}{2})$,
which is the direct sum of two $SU(2)$ gauge theory.  Thus, the
$SU(2)$ Berry phase naturally arises in this system.

This $SU(2)$ nature of the Berry phase in the spin-3/2 system is
manifested if we choose a special spinor $|\alpha\rangle$ such
that ${\bf \Delta}_{\bf k} |\alpha\rangle  =|\Delta_{\bf
k}|e^{-i\phi_\alpha}|\alpha\rangle$, which is studied by Demler
and Zhang\cite{DEMLER1999}. In this representation,
$|\alpha\rangle$ is not only a $SU(4)$ spinors but also a $SO(5)$
one. Then, the BPC is given by
\begin{eqnarray}
  A^{\pm \alpha\beta}_i=\langle \psi^\pm_\alpha|\partial_i|\psi^\pm_\beta\rangle
  =\langle \alpha|\partial_i|\beta\rangle.
\end{eqnarray}
This is exactly the $SU(2)$ holonomy in the context of Demler and
Zhang\cite{DEMLER1999}.  The special choice of the spinors
$|\alpha\rangle$ is equivalent to fixing $\hat{d}_a=\hat{d}_5$ in
our notion.

\section{Conclusion and discussion}\label{SECTION-Conclusion}

In summary, we found that the SU(2) non-Abelian Berry phase
emerges naturally in quintet condensates of spin 3/2 fermions. The
underlying algebraic structures for the $^3He$ and the spin-3/2
system are the $1^{st}$ and the $2^{nd}$ Hopf map respectively.
The Chern numbers for both cases were obtained in a standard
manner.  In the previous case, only the first Chern number is
non-vanishing, while in the later case, only second Chern number
is non-zero.  Both systems appear to have \emph{finite} gauge
potential, which means that the BPC can be defined covariantly
everywhere over the $d$-field.  However, the corresponding $U(1)$
and $SU(2)$ gauge connections can only be defined patch by patch
in the $d$-space. The bridge across between the finite gauge
connection and the one with singularity is constructed by the
singular gauge transformation, which can only be defined patch by
patch in the $d$-space as well\cite{Moody1986}.  The singular gauge
transformation also bear with some physical meaning.  It can be
understood as the rotation in the spin (isospin) space, namely
$d_a$.  When the spin (isospin) points to the north pole, the
invariant subgroup becomes the isometry group of the equator. When
it is rotated by the gauge transformation, the gauge structure
becomes finite and covariant over the whole $d$ space.

To experimentally manifest this topological effect, the spin-3/2
ultra-cold atomic systems may serve as a promising candidate.  It
may also shed some light on measuring the second Chern number,
which has not be revealed by any system so far.  Furthermore, our
calculation can also be generalized to consider the
spin-$\frac{7}{2}$ superconductors in which the algebraic
structure is suggested to be the third Hopf
map\cite{Bernevig2003}.

This work is supported by the NSF under grant numbers DMR-0342832
and the US Department of Energy, Office of Basic Energy Sciences
under contract DE-AC03-76SF00515.  HDC and CJW are also supported
by the Stanford Graduate Fellowship.


\end{document}